\documentclass[runningheads]{llncs}
\usepackage[T1]{fontenc}
\usepackage{graphicx}

\usepackage{subcaption}
\usepackage{caption}
\usepackage{booktabs} 
\usepackage{listings}
\usepackage{xcolor} 
\usepackage{amsmath}
\usepackage{amssymb}
\usepackage{multirow}

\begin{document}
\title{FAIR-QR: Enhancing Fairness-aware Information Retrieval through Query Refinement}
\titlerunning{Enhancing Fairness-aware Information Retrieval through Query Refinement}
%

\author{Fumian Chen\inst{1}\orcidID{0009-0001-2391-6578} \and
Hui Fang\inst{1,2}\orcidID{0009-0003-1904-787X}\\
\institute{Institute for Financial Services Analytics, University of Delaware, USA
\and Department of Electrical and Computer Engineering, University of Delaware, USA}
\email{\{fmchen,hfang\}@udel.edu}}
\authorrunning{F. Chen and H. Fang}
%
%
\maketitle              
\begin{abstract}
Information retrieval systems such as open web search and recommendation systems are ubiquitous and significantly impact how people receive and consume online information. Previous research has shown the importance of fairness in information retrieval systems to combat the issue of echo chambers and mitigate the rich-get-richer effect. Therefore, various fairness-aware information retrieval methods have been proposed. Score-based fairness-aware information retrieval algorithms, focusing on statistical parity, are interpretable but could be mathematically infeasible and lack generalizability. In contrast, learning-to-rank-based fairness-aware information retrieval algorithms using fairness-aware loss functions demonstrate strong performance but lack interpretability. In this study, we proposed a novel and interpretable framework that recursively refines query keywords to retrieve documents from underrepresented groups and achieve group fairness. Retrieved documents using refined queries will be re-ranked to ensure relevance. Our method not only shows promising retrieval results regarding relevance and fairness but also preserves interpretability by showing refined keywords used at each iteration.

\end{abstract}

\section{Introduction}

In the current information explosion era, information retrieval (IR) systems are essential in how people receive and consume information. For example, a search engine might retrieve biased information that only covers information from one perspective, which might cultivate an echo chamber and even polarize the online community as more and more people rely on search engines for information needs \cite{patro2022fair,dai2024bias}. To address these issues, many fairness-aware information retrieval algorithms have been proposed \cite{zehlike2022fairness}. For example, score-based methods enforce minimal exposure of protected groups on the search engine result page (SERP) and are interpretable by directly bringing more exposure to protected groups. However, they often require tailored calibration and are even mathematically impossible \cite{zehlike2022fairness}. Even though some methods use greedy algorithms, global solutions are not guaranteed \cite{gao2022fair}. In contrast, learning-to-rank-based methods have shown promising results utilizing fairness-aware training losses or other fairness-aware constraints \cite{zehlike2022fairness}. Nevertheless, their black-box mechanisms make them less interpretable, let alone supervised learning-based methods that require ground truth labels that are barely available \cite{chen2023learn}. Given the increasing attention to fairness-aware IR systems and existing work's limitations, an interpretable, generalizable, and well-performed fairness-aware IR algorithm is pressingly needed, especially when more and more people are gravitating to IR-based applications such as retrieval augmented generation (RAG) systems, but their retrievers' fairness is overlooked \cite{dai2024bias,kim2024towards}.

Query refinement (QR) is a technique to improve the effectiveness of IR systems by changing or modifying queries to retrieve more user-desired results \cite{guo2008unified}. The most well-applied QR approach uses relevance feedback or search history to improve the accuracy of search queries \cite{ooi2015survey}. With the rapid development of natural language processing (NLP), query refinement is moving to the next level. Advanced NLP techniques, like language models, are proven to capture contextual information accurately \cite{guo2008unified}. Models like Generative Pre-trained Transformer (GPT) use self-attention mechanisms to capture the semantic relationships between words across different positions in a text sequence \cite{bengesi2024advancements}. This allows the models to generate coherent, well-linguistically structured, relevant responses, which are the bedrock to refine queries using these models. Since search query is one the most determinant factors of SERPs, it also impacts the fairness of search results. One of the previous studies shows that explicitly including different gender terms in query keywords produces a significant difference in gender representation in the search results \cite{kopeinik2023show}. Another study de-biases query keywords toward gender using a pseudo-relevance feedback approach \cite{bigdeli2021orthogonality}, but it focuses on binary fairness groups and does not comply with the well-adapted exposure-based fairness definitions, where fair exposures of different groups are not always equal but proportional to their relevance. Therefore, refining queries to achieve fairness has great potential but is under-exploited.  

This work proposed a QR-based method to recursively bring documents from under-exposed groups to SERP. We use a divergence-based fairness constraint to control the number of iterations and the generation quality. As we modify queries, it is possible that relevance is compromised. Therefore, we semantically re-rank the retrieved documents to minimize relevance loss. Compared with existing score-based or learning-to-rank-based fair ranking algorithms, our method focuses on QR. It is more interpretable than learning to rank methods since refined queries are trackable at each step. It is also more generalizable than score-based methods because it does not require specific calibrations. To the best of our knowledge, we proposed the first framework using query refinement to achieve fairness-aware IR. Our source code and results are publicly available \footnote{\url{https://github.com/fm-chen/FAIR-QR}}.


\begin{figure}[t]

    \centering
    \includegraphics[width=1\linewidth]{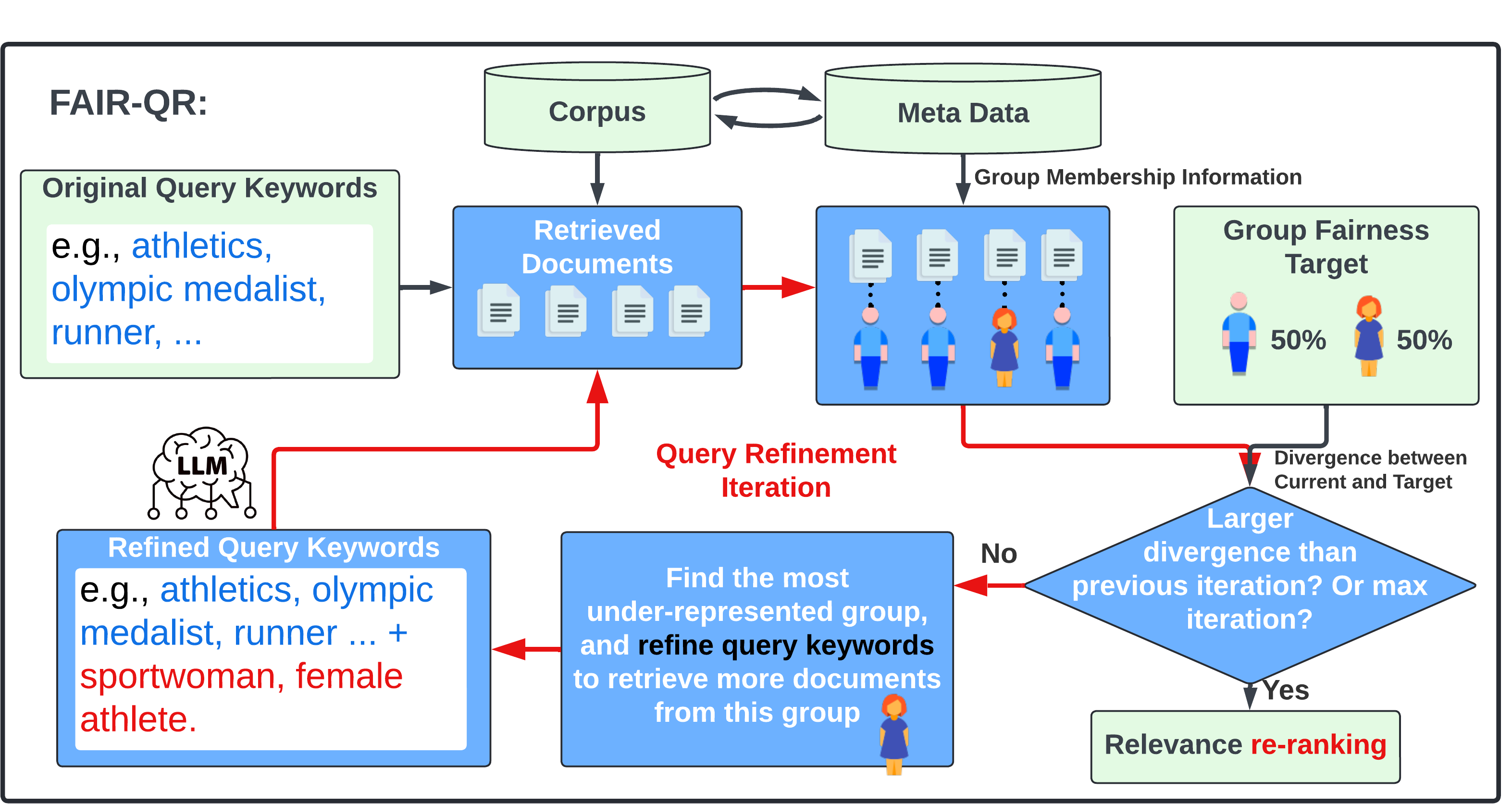}
    \caption{FAIR-QR workflow. Refined query keywords are visible at each iteration.}
    \label{fig:genQR-flow}
    
\end{figure}

\section{Methodology}

\subsection{Problem Formulation}

Given a corpus $D$ and a set of query $Q$, information retrieval systems retrieve a set of ranked documents $\mathcal{L}_q = (d_1, d_2, ..., d_n)$ where $d_i \in D$ for each query $q \in Q$. For each document retrieved, group memberships (GM) $g \in \mathcal{G}$ are assigned or annotated to show the document's meta information, such as the geographic location and gender of the authors associated with the document. Given a query to retrieve a set of ranked documents, fairness-aware IR systems (1) rank \textbf{relevant} documents in higher positions and (2) satisfy our \textbf{fairness} definition. 

Various fairness definitions have been defined based on different fairness goals. This work adopts the well-accepted exposure-based group fairness definition. That is, \textit{no group should receive more or less expected exposure than groups with the same relevance} \cite{diaz2020evaluating}. Given a fairness category (e.g., gender), the fairness goal is to make sure that each subgroup (e.g., male, female, non-binary) receives expected exposure $\epsilon^*$.

\subsection{FAIR-RQ} \label{fair-genqr}

Fig.\ref{fig:genQR-flow} shows the workflow of fairness-aware information retrieval by query refinement (FAIR-QR) we proposed. It consists of three major tasks: document retrieval, recursively query refinement, and relevance-based re-ranking. The retrieval task, a well-studied field, aims to retrieve relevant documents based on given queries. FAIR-RQ focuses on the query refinement task.


One pre-requisite component of query refinement is each document's group membership information $g^{d_i} \in \{1,0\}$ to formulate the current exposure distribution of a retrieved document list $\epsilon_{\mathcal{L}} = (g_1, g_2, ..., g_m)$ for $m$ groups where $g_i = \sum_{1}^n \frac{g^{d_i}}{n}$. For example, if 70\% of retrieved documents are from the male group, whereas 30\% of documents are from the female group, then, $\epsilon_{\mathcal{L}} = (g_\text{male}, g_\text{female}) = (0.7, 0.3)$. We compare the current exposure distribution $\epsilon$ with a target exposure distribution $\epsilon^*$. If the difference (quantified by KL divergence\footnote{\url{https://en.wikipedia.org/wiki/Kullback-Leibler\_divergence}}) between these two distributions $\Delta(\epsilon, \epsilon^*)$ is not small enough, some groups are underrepresented than expected, and we keep refining queries to retrieve more documents from the underrepresented groups. Detailed steps are: (1) we identify the most underrepresented group $\hat{g}$ in the retrieved documents. (2) we refine the query using LLMs because $\hat{g}$ is underrepresented, and related keywords should be added or modified to increase the exposure of $\hat{g}$ in the retrieved documents. We use one of the most powerful LLMs, the GPT-4o model (we set temperature = 0.3 using a grid search approach for the best performance and re-productivity) \cite{openai2023gpt4}. FAIR-RQ can incorporate different LLMs, too. (3) go to the retrieval task to retrieve documents using refined queries. (4) repeat (1)-(3) until $\Delta(\epsilon, \epsilon^*)$ does not decrease or max iteration. 

For the query refinement method, an intuitive way is to add keywords related to the underrepresented group. Existing query refinement algorithms focus on relevance and are not designed for fairness-aware IR. This study uses generative models to refine queries since they have been proven to accurately generate relevant content in response to prompts. To address its difficulty in controlling model output, we utilize a chain-of-thought mechanism to refine the query recursively. The best prompt we use, shown in the Appendix, is inspired by previous studies \cite{sahoo2024systematic} and obtained by some ablation experiments. Given the space limit, more details will be shared in our source code repository.

Lastly, after refining our query to bring more documents from underrepresented groups, relevance might be jeopardized as we add or modify query keywords. Hence, we re-rank the retrieved documents using the original query to ensure that the most relevant documents are ranked at the top. The re-ranking function is the same retrieval function used in the retrieval task and, ideally, could be any retrieval function. We use the simple yet effective BM25 \cite{robertson1995okapi}.


\section{Experiments}

\begin{table}[!t] 

\centering
\caption{Model relevance and fairness performance comparison (nDCG*AWRF is reported to show the balance between fairness and relevance). $\divideontimes$,$\diamond$,$\P$,$\S$,$\dag$, and $\ddag$ indicate statistically significant better performances of \textit{FAIR-QR} than \textit{BM25($\divideontimes$), Vector Search($\diamond$), DLF($\P$), DELTR($\S$), FA$\ast$IR($\dag$), MMR($\ddag$)}, respectively (paired t-test based on 47 evaluation queries with p-value<0.05).}
\label{tab:fair_genqr}
\resizebox{\textwidth}{!}{%
\begin{tabular}{c|c|cccc}
\hline \hline 
\multirow{2}{*}{} &
  \multirow{2}{*}{\textbf{nDCG@20$\uparrow$}} &
  \multicolumn{2}{c|}{\textbf{Biographic Gender}} &
  \multicolumn{2}{c}{\textbf{Geographic Location}} \\ \cline{3-6} 
      &        & \multicolumn{1}{c|}{AWRF@20$\uparrow$} & \multicolumn{1}{c|}{nDCG*AWRF$\uparrow$} & \multicolumn{1}{c|}{AWRF@20$\uparrow$} & nDCG*AWRF$\uparrow$ \\ \hline \hline 
BM25 Retrieval \cite{robertson1995okapi} &
  0.6638 &
  \multicolumn{1}{c|}{0.8857} &
  \multicolumn{1}{c|}{0.5879} &
  \multicolumn{1}{c|}{0.7895} &
  0.5241
  \\ 
  Vector (SBERT) Search \cite{reimers-2019-sentence-bert} &
  0.4771 &
  \multicolumn{1}{c|}{0.8546} &
  \multicolumn{1}{c|}{0.4233} &
  \multicolumn{1}{c|}{0.7560} &
  0.3619
  \\\hline 
FAIR-QR [no re-rank] &
  0.6423 $\diamond$ &
  \multicolumn{1}{c|}{0.9296$\divideontimes$$\diamond$$\P$$\S$$\dag$$\ddag$} &
  \multicolumn{1}{c|}{0.5971$\diamond$$\S$} &
  \multicolumn{1}{c|}{\textbf{0.8326$\divideontimes$$\diamond$$\P$$\S$$\dag$$\ddag$}} &
  0.5348$\diamond$$\S$ \\
  
  \textbf{FAIR-QR} &
  0.6530 $\diamond$ &
  \multicolumn{1}{c|}{\textbf{0.9316$\divideontimes$$\diamond$$\P$$\S$$\dag$$\ddag$}} &
  \multicolumn{1}{c|}{\textbf{0.6083$\diamond$$\P$$\S$$\dag$}} &
  \multicolumn{1}{c|}{0.8314$\divideontimes$$\diamond$$\P$$\S$$\dag$$\ddag$} &
  \textbf{0.5429$\divideontimes$$\diamond$$\P$$\S$$\dag$} \\
  \hline 
DLF \cite{chen2023learn}   & 0.6512 & \multicolumn{1}{c|}{0.9013}  & \multicolumn{1}{c|}{0.5869}    & \multicolumn{1}{c|}{0.8108}  & 0.5280    \\
DELTR \cite{zehlike2020reducing} & 0.6524 & \multicolumn{1}{c|}{0.8850}  & \multicolumn{1}{c|}{0.5774}    & \multicolumn{1}{c|}{0.7786}  & 0.5079    \\ \hline 
FA$\ast$IR \cite{zehlike2017fa} & 0.6603 & \multicolumn{1}{c|}{0.8862}  & \multicolumn{1}{c|}{0.5852}    & \multicolumn{1}{c|}{0.7950}  & 0.5249    \\
MMR \cite{carbonell1998use}  & \textbf{0.6686} & \multicolumn{1}{c|}{0.8795}  & \multicolumn{1}{c|}{0.5880}    & \multicolumn{1}{c|}{0.7960}  & 0.5321    \\ \hline \hline 
\end{tabular}%
}

\end{table}

\subsection{Experiment Settings}

\subsubsection{Experimental Dataset.} We select the most recent fair ranking dataset, the TREC fair ranking track 2022 \cite{ekstrand2023overview}, as our experimental dataset to train and evaluate fair ranking models. The corpus comprises more than six million English Wikipedia articles with full-text fields. Each article is also associated with its group membership information (e.g., geographic location and biographic gender). We focus on retrieving Wikipedia articles for Wikipedia editors and ensuring the results are relevant and fair. 47 evaluation queries from various domains and their ground truth relevant documents are provided.

\subsubsection{Fair Ranking Baselines.} Our baseline models include (1) a relevance-based retrieval model, BM25 \cite{robertson1995okapi}, implemented by Elasticsearch \footnote{\url{https://www.elastic.co/elasticsearch}} (k1 = 1.2 and b = 0.75), (2) a vector search model using \textit{Sentence-Transformer} embeddings \cite{reimers-2019-sentence-bert} and cosine similarity with its pre-trained model \textit{'all-mpnet-base-v2'}, (3) a recent learning to rank model using distribution-based fairness loss, DLF ($\alpha=0.8$) \cite{chen2023learn}, (4) another learning to rank model using fairness-aware loss, DELTR ($\gamma=500$) \cite{zehlike2020reducing}, (5) a score-based post-processing method, FA$\ast$IR ($k=400$, $p=0.3$, $\alpha=0.15$) \cite{zehlike2017fa}, and (6) a diversification-based model, Maximal Marginal Relevance (MMR) ($\lambda=0.5$) \cite{carbonell1998use}.

\subsection{Results and Analysis}

We evaluate algorithms based on their performance in terms of both relevance and fairness. We choose nDCG\footnote{\url{https://en.wikipedia.org/wiki/Discounted\_cumulative\_gain}} as relevance evaluation. Given our group fairness goal, we use the same fairness evaluation metric as the TREC fair ranking track, the attention-weighted ranking fairness (AWRF) \cite{sapiezynski2019quantifying,raj2022measuring}: $\text{AWRF}(\mathcal{L}) = 1 - \Delta(\epsilon, \epsilon^*)$ where $\epsilon^*$ is a fair ranking target exposure distribution, $\epsilon$ is a system produced exposure distribution, and $\Delta$ is the Jenson-Shannon divergence. The target exposure distribution $\epsilon^*$ is based on the empirical distribution of groups among all relevant documents. We evaluate fairness regarding biography gender (4 subgroups including male, female, non-binary, and unknown) and geographic location (21 subgroups, including Asia, North America, etc.).

According to Section \ref{fair-genqr} and the workflow shown in Fig.\ref{fig:genQR-flow}, the maximal number of iterations needs to be set for optimal performance. Since iterations will be terminated earlier if $\Delta(\epsilon, \epsilon^*)$ does not decrease, we set the maximal number of iterations very high (20) at first and examine how many iterations were proceeded for the 47 evaluation queries. For both biography gender and geographic location, we found that the max number of iterations is 11, the mean number of iterations is 2, and the 75th quintile is 4. Therefore, we set the maximal number of iterations to 5 for the best generalizability and to avoid over-fitting. Table \ref{tab:fair_genqr} reports the relevance and fairness performance of FAIR-QR compared with other baseline models. We notice that the vector search method performed much worse than BM25, which is unsurprising since our query was keywords extracted from relevant documents with KeyBERT \cite{grootendorst2020keybert} other than query sentences with sequential meanings \cite{ekstrand2023overview}. 

\begin{figure}[!t]

    \centering
    \begin{subfigure}{0.5\textwidth}
        \centering
        \includegraphics[width=\linewidth]{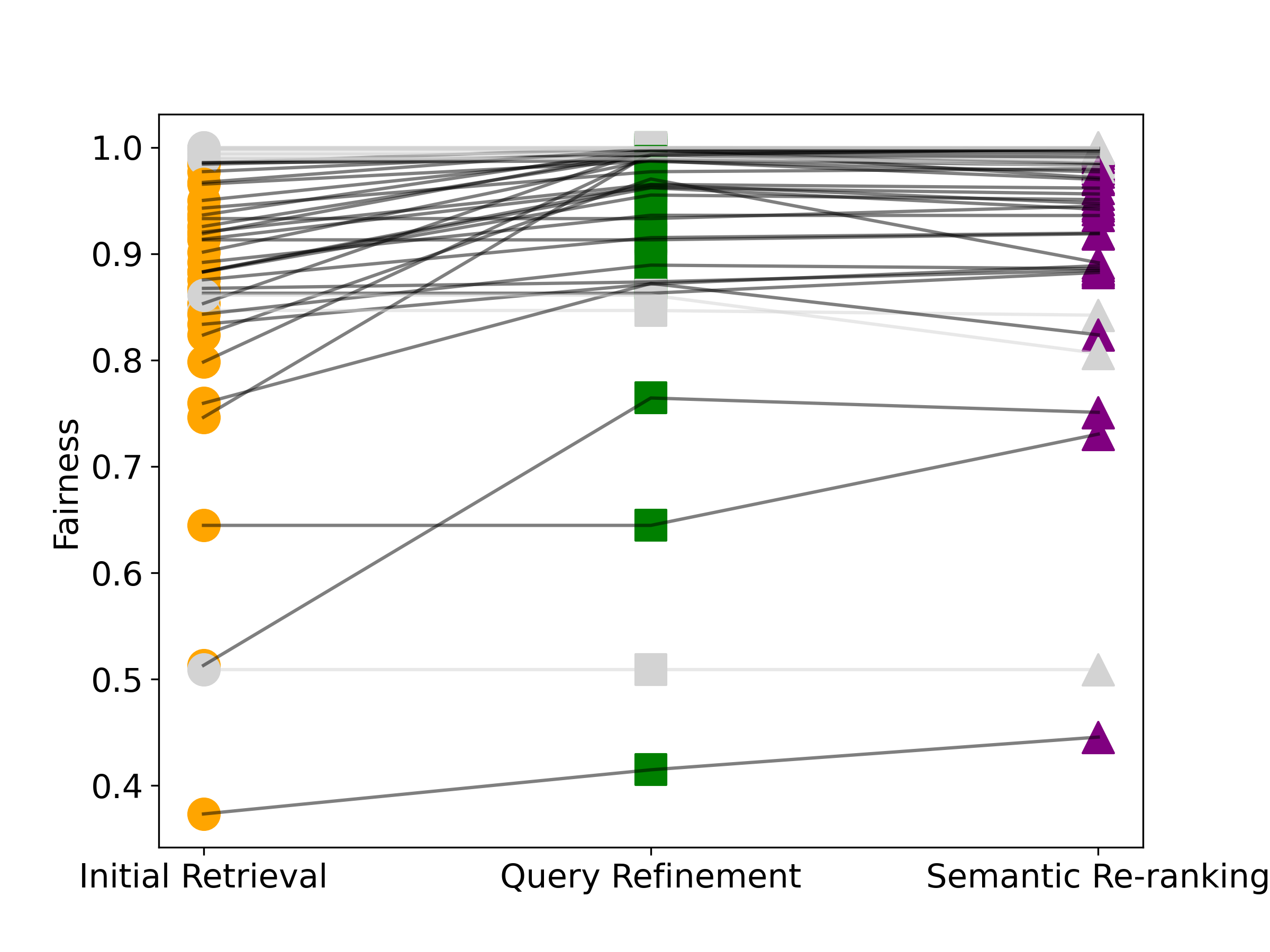}
        \caption{Biography Gender}
        
    \end{subfigure}\hfill
    \begin{subfigure}{0.5\textwidth}
        \centering
        \includegraphics[width=\linewidth]{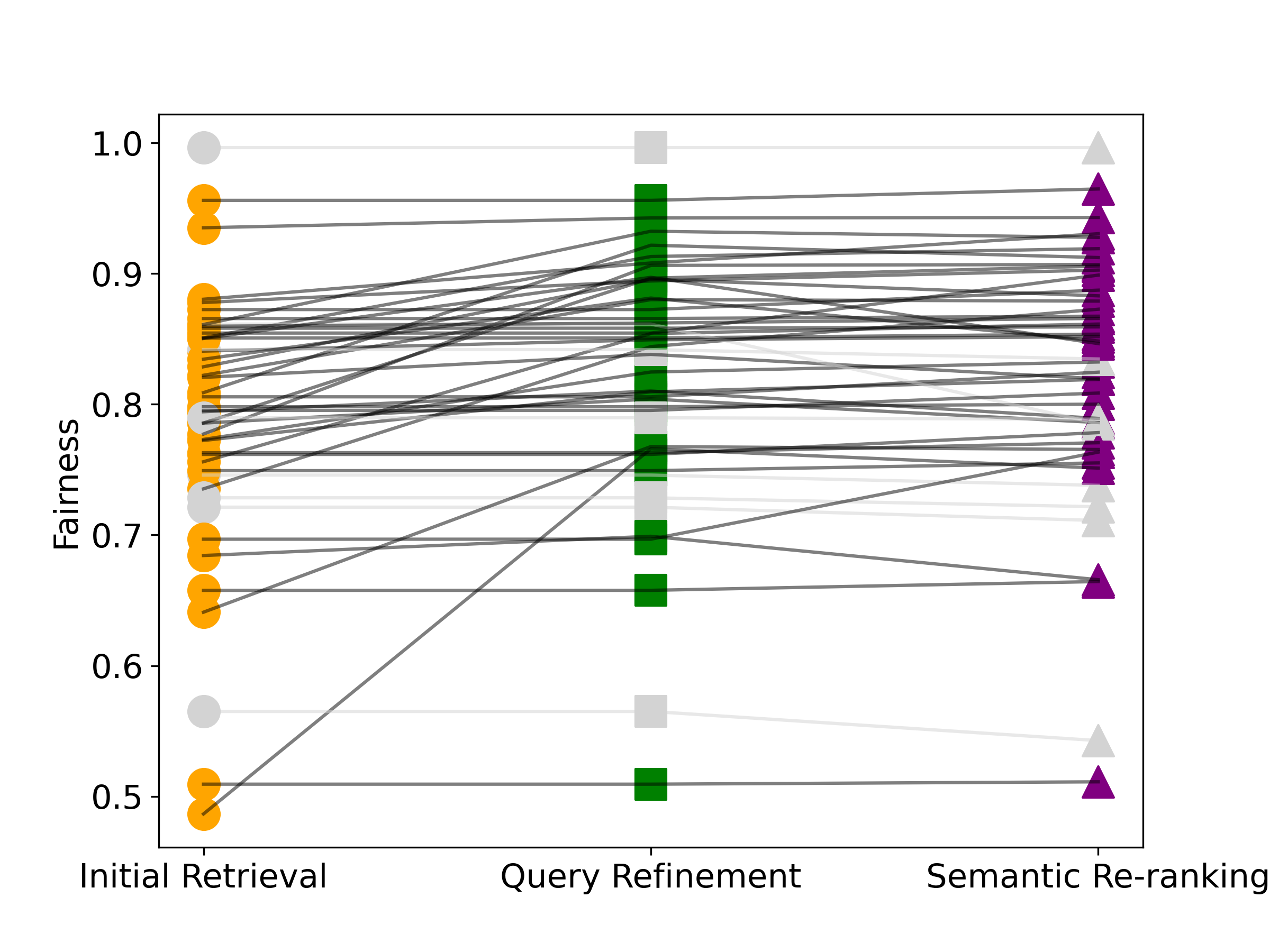}
        \caption{Geographic Location}  
        
    \end{subfigure}
    \caption{Fairness improvements with query refinement and semantic re-ranking based on 47 evaluation queries (Solid shapes/lines indicate improvements).}
    \label{fig:stage_improv}
\end{figure}

As can be seen from the table, FAIR-QR outperformed all baselines regarding the fairness performance of biographic gender and geographic location. Even though FAIR-QR's relevance performance is not the best, it achieved the highest score of nDCG$*$AWRF, reflecting a balanced fairness and relevance performance, especially with the relevance-based re-ranking. The diversification-based MMR improved relevance but not fairness, showing the complexity of balancing fairness and relevance in this task. This demonstrates the effectiveness of FAIR-QR in improving ranking fairness using generative query refinement with a minimal loss of relevance. When comparing the results between gender and geographic location, we notice that fair ranking is more challenging regarding fairness categories with more subgroups. FAIR-QR can improve fairness constantly regardless of fairness categories and better manage the relevance-fairness trade-off.

Fig. \ref{fig:stage_improv} demonstrates how fairness was improved in each stage of FAIR-QR about the 47 evaluation queries. We noticed that fairness improved during the query refinement and was maintained after semantic re-ranking for most queries regarding gender and geographic location. This shows the effectiveness of FAIR-QR in producing fair results for queries from various domains. Compared with the relevance performance reported in Table \ref{tab:fair_genqr}, the semantic re-ranking stage balanced fairness and relevance effectively.

\section{Conclusion}

In this study, we proposed a novel fairness-aware IR algorithm, FAIR-QR, by recursively refining queries. Compared with existing fairness-aware IR algorithms, FAIR-QR not only achieved better fairness performance but also balanced relevance through relevance re-ranking. FAIR-QR is more interpretable than black-box methods since query keywords are tracked during query refinement, as demonstrated in Fig. \ref{fig:genQR-flow}. Ideally, FAIR-QR can be applied to any retrieval algorithms and various generative models. We tested BM25 and GPT-4o in this work, and more different models will be explored in future studies. Furthermore, since the data sparsity, we only tested FAIR-QR on one recent fair-ranking dataset, but it can be applied to general datasets with group membership annotations. On the other hand, we showed the best prompt inspired by previous studies that gave us the best results. Given the focus of this work, we leave further prompt engineering to future work. Even though this is still an exploratory study and there are spaces for improvements, it opens up a promising direction for utilizing generative AI to improve fairness in IR systems by query refinement.

\begin{credits}
\subsubsection{\ackname} The research was supported as part of the Center for Plastics Innovation, an Energy Frontier Research Center, funded by the U.S.\ Department of Energy (DOE), Office of Science, Basic Energy Sciences (BES), under Award Number DE-SC0021166. It was also supported by the IFSA at the University of Delaware. The authors thank the reviewers for their constructive comments and suggestions.
\end{credits}

\vspace{-3mm}
\section*{Appendix: Prompt Used for Query Refinement}
\vspace{-3mm}
\begin{lstlisting}
You are a user who cares about the fairness of a search engine where searched documents are retrieved from different subgroups. 
You want to make sure the retrieved documents of query: {Query} are from diverse fairness groups quantified by a target distribution: {Target Exposure Distribution}, which shows the desired percentage of retrieved documents from each subgroup. The keys in the target distribution are the unique subgroups. The 'Unknown' subgroup means group information is missing or not applicable.
Now, using the BM25 method, you got results of the first {Top_K} documents with a fairness group distribution of: {Current Exposure Distribution}. You want to achieve the target by adding keywords or phrase at the end of the original query with less jeopardize relevance. Therefore, you must add less keywords as possible to make the current results more align with our fairness target distribution and remain relevant. 
Let's try to focus on the subgroup that is most under-represented. In this case, it's the subgroup: {subgroup}. Show me your refined keywords that can help retrieve a composation of documents from different gender group closer to the target distribution. That is, knowing the retrieved documents have fewer than desired documents from group {subgroup}, you might want to include keywords about {subgroup}.    
\end{lstlisting}

%
%
%
\bibliographystyle{splncs04}
\bibliography{my}

\end{document}